# Regulation of dispersion of carbon nanotubes in a mixture of good and bad solvents

O. Deriabina[a,b], N. Lebovka[a]*, L. Bulavin[b], A. Goncharuk[a]

[a]*Institute of Biocolloidal Chemistry named after F.D. Ovcharenko, NAS of Ukraine, 42, blvr. Vernadskogo, Kyiv 03142, Ukraine*

[b]*Taras Shevchenko Kiev National University, Department of Physics, 2, av. Academician Glushkov, Kyiv 031127, Ukraine*

**Abstract**

The microstructure and electrical conductivity of suspensions of multi-walled carbon nanotubes (MWCNTs) in binary liquid mixtures water + 1-Cyclohexyl-2-pyrrolidone (CHP) were studied in the temperature interval of 253–318 K, in the heating and cooling cycles. The concentration of MWCNTs was varied in the interval between 0-1 wt. % and the content of water in a binary mixture $X$ = [water]/( [CHP] + [water]) was varied within 0-1.0. The experimental data have shown that dispersing quality of MWCNTs in a mixture of good (CHP) and bad (water) solvents may be finely regulated by adjustment of composition of the CHP+ water mixtures. The aggregation ability of MWCNTs in dependence on X was discussed. The surface of MWCNT clusters was highly tortuous, its fractal dimension $d_f$ increased with increase of $X$, approaching $\approx 1.9$ at $X \to 1$. It was concluded that the surface tension is not suitable characteristic for prediction of dispersion ability in the mixture of good and bad solvents. The electrical conductivity data evidenced the presence of a fuzzy-type percolation with multiple thresholds in the systems under investigation. This behavior was explained by formation of different percolation networks in dependence of MWCNT concentration.

## 1. Introduction

Nowadays, materials, based on carbon nanotubes (MWCNTs) demonstrate many attractive electrical, mechanical and thermal properties with potential of their applications in engineering, electronics and medicine. The different types of electrochemical biosensors, piezoresistive strain sensors, electromagnetic switches, screens devices, electrical energy and memory storage devices, supercapacitors, multifunctional polymer nanocomposites, materials for heat transfer devices, ion battery, ultrafast photonics, sorbents, biopharmaceutics and tissue engineering were already proposed [1–4].

Poor dispersability of MWCNTs in many solvents and tendency for formation of the bundles or large aggregates presents a serious obstacle to good functionality of MWCNT-based materials. E.g., the quality of MWCNT dispersion in suspensions and derived functional composites may be particularly important in determination of electrical, mechanical and thermal properties of these systems. The studies have shown that shearing processing strongly affected the percolation behavior and clustering of MWCNTs in different composites [5–7].

Colloidal processing is a recognized tool for preparation of MWCNT-based materials[8,9] and regulation of their stability [10–13]. In some cases, colloidal suspension of nanotubes may serve as a useful intermediate for further processing and preparation of more complex nanotube-based composites. Because of high hydrophobicity of MWCNT surface, water is a bad solvent for MWCNTs. Colloid stability of aqueous suspensions may be improved by oxidation or functionalization of MWCNTs[14–16] and introduction of different surfactants [10–12,17–20], polymers

*Corresponding author. Tel: +380(44)4240378, Fax: +380(44)4248078. *E-mail address:* lebovka@gmail.com (N.I. Lebovka).

[21] or supplementary colloidal particles [22–25]. Note that the dispersability of MWCNTs may be noticeably improved in some non-aqueous solvents or their mixtures [26–32]. MWCNTs can be debundled to a significant degree in such a good solvent as N,N-dimethylformamide (DMF), N-methyl-2-pyrrolidone (NMP), 1-Cyclohexyl-2-pyrrolidone (CHP) [26,32]. E.g., the aprotic CHP can disperse MWCNTs up to 3.5 mg/mL with very large populations of individual MWCNTs[29]. It was suggested[28,33,34] that good dispersion ability of solvent is determined by the values of their Hansen solubility parameters [35]. It was also speculated that good MWCNT-dispersing ability of solvents is due to the low energetic cost of exfoliation of nanotubes and because the surface tensions of solvents successfully dispersing MWCNTs are close to 40 mJ/m$^2$ [29]. Application of these arguments for prediction of MWCNT-dispersing ability of a mixture of solvents is still unclear. So, it is interesting to study the dispersion efficiency of a mixture of good and bad solvents. A suitable candidate is the mixture of CHP and water that have significantly different total Hansen parameters: $\delta_t$=47.8 MPa$^{1/2}$ and 20.5 MPa$^{1/2}$, respectively [35]. This difference mainly reflects the differences in dispersion and hydrogen bonding components of Hansen parameters. A remarkable property of CHP-water mixture is that CHP is fully miscible with water below their lower consolute temperature (LCT) $T \approx 320$ K,. and above this temperature the liquids form two separate phases [36,37].

This work was devoted to the study of microstructure of the colloidal suspensions of MWCNTs in CHP + water mixtures at different concentrations of solvent components. The temperature dependences of electrical conductivity at different concentrations of MWCNTs and percolation behavior of the systems were also investigated.

## 2. Experimental

*2.1 Materials*

The multi-walled carbon nanotubes (MWCNTs) were prepared from ethylene using the chemical vapor deposition method (TMSpetsmash Ltd., Kyiv, Ukraine) with Fe–Al–Mo catalyst [38]. MWCNTs were further purified by alkaline and acidic solutions and washed by distilled water until reaching the distilled water pH and conductivity values in the filtrate. The typical outer diameter of MWCNTs $d_n$ estimated from electrical microscopy images, was 20 – 40 nm[39], while their length $L_n$ ranged from 5 to 10 μm. The specific surface area of the MWCNT powder $S_n$, determined by $N_2$ adsorption, was 130 ± 5 m$^2$/g. The specific electric conductivity $\sigma_p$ of the powder of MWCNTs compressed at 15 TPa was about 10 S/cm along the axis of compression.

1-Cyclohexyl-2-pyrrolidone (CHP) with 99.5% purity (Sigma Aldrich Co., USA) was used as a good solvent. It is a high-boiling (boiling point is 427 K/7 mmHg) aprotic solvent with molecular formula $C_{10}H_{17}NO$ (Fig.1) and density (1.007 g/mL at 298 K) close to that of water. The melting point of CHP is 285 K. Note that CHP is a surface active substance with relatively large critical micelle concentration in water (CMC≈0.45 M) [36]. The distilled but not degassed water with conductivity of $\sigma_o$≈80 μS/cm (measured at 298 K and the frequency of 10$^3$ Hz) was used as a bad solvent.

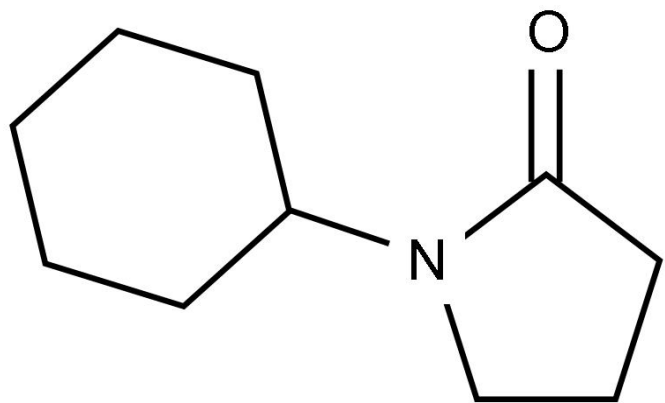

Fig. 1. Chemical structure of CHP.

The CHP + water mixtures were prepared by adding the appropriate weights of CHP into water. The weight concentration of water in CHP was expressed as $X =$ [water]/([CHP]+ [water]). To obtain a CHP + water + MWCNT suspension, the necessary portion of MWCNTs was added into CHP + water mixture. Then suspension was sonicated using the ultrasonic disperser UZDN-2T (Ukrrospribor, Ukraine) at the frequency of 44 kHz and the power of 400 W. To prevent suspensions from overheating, they were sonicated in a cold-water bath, the temperature of suspension never exceeded 313 K. Sonication promotes dispersion of highly entangled or aggregated MWCNT samples, but also results in the damage of MWCNTs [22]. In our experiments, each suspension was sonicated for 10 min and used immediately after preparation.

The series of samples with concentration of MWCNTs $C$ within 0.0 – 1.0 % wt and water concentration $X$ within 0 – 1 were investigated.

*2.2 Instrumentation*

The optical microscopy images of suspensions were obtained using the optical microscope Biolar (03-808 Warszawa, Poland). During the measurements, suspension was placed in a dispersion cell; the thickness of suspension layer was 130 μm. The microscope was equipped with a digital camera, connected to a personal computer.

The black and white two-dimensional projections of images were used for analysis of the microstructure of composites. The cluster analysis was performed using Hoshen-Kopelman algorithm [40], then mean radius of aggregates $<r>$ and their size distribution $F(r)$ were determined. The box-counting method and the image analysis software ImageJ v1.42q were applied for estimation of the 'capacity' of fractal dimension $d_f$. The value of $d_f$ was obtained from dependence of the number of boxes, necessary for covering the boundary of an aggregate, $N$, versus the box size, $s$:

$$N \propto s^{d_f} \qquad (1)$$

The electrical conductivity measurements were carried out in the temperature interval of 253–318 K, in the heating and cooling cycles, with a constant rate of scanning 2 K min$^{-1}$. The temperature was stabilized using a flowing-water thermostat Type 100/45 110 (Mechanik Pruefgeraete Medingen, Germany), and was recorded by a Teflon-coated K-type thermocouple (±0.1 K), connected to the data logger thermometer centre 309 (JDC Electronic SA, Switzerland). The cell for electrical conductivity measurement included two horizontal platinum electrodes of 13.5 mm in diameter, with 0.5 mm inter-electrode space. Before the measurements, the cell parts were washed in hexane and dried at 390 K. The electrical conductivity of samples was measured by AC bridge AM 303 (Data.Com, Russia). The measurements were made at the frequency of $10^3$ Hz. This frequency was selected for avoidance of polarization effects on the electrodes and electrical field-induced asymmetric redistribution of MWCNTs between the electrodes [41].

The preliminary studies revealed a noticeable hysteresis of electrical conductivity $\sigma$ in the heating and cooling cycles. Figure 2 present examples of $\sigma$ temperature dependences in three consecutive cycles at the same concentration of MWCNTs, $C=0.3$ % wt, a) in pure CHP ($X=0$) and b) CHP + water mixture ($X=0.15$). The similar hysteresis behavior was observed at other values of $X$. This phenomenon may reflect the restructuring of MWCNT aggregates during the heating-cooling cycles. The hysteresis loops were becoming smaller with increase of the number of cycles, so, for certainty of results, all the measurement data were presented only for the third heating-cooling cycle.

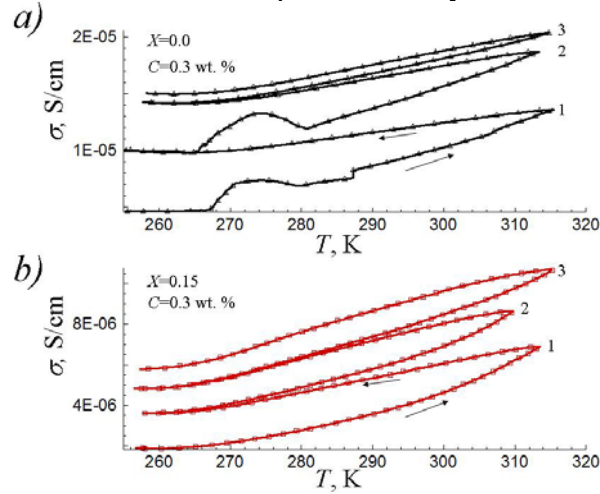

Fig. 2. Examples of temperature dependences of electrical conductivity in MWCNT suspensions at: a) $X=0.0$ (pure CHP) and b) $X=0.15$ at $C=0.3$ wt. %. The numbers 1, 2, 3 correspond to the consecutive heating-cooling cycles.

*2.3 Statistical analysis*

Each measurement was repeated, at least, three times for calculation of the mean values and root-mean-square errors. The error bars in all the figures correspond to the confidence level 95 %. The least square fitting of the experimental dependencies, determination of the fitting parameters and correlation coefficient $r^2$, were provided using Table Curve 2D software (Jandel Scientific, USA).

## 3. Results and discussion

Figure 3 shows examples of micro-photos of MWCNT suspensions in CHP + water mixtures at different values of $X =$[water]/([CHP]+ [water]) and fixed value of $C=0.1$ wt. %.

The dispersions of MWCNTs were almost ideal and the MWCNT aggregates were not detected visually in a good solvent (pure CHP). It means that such suspensions had very large populations of individual MWCNTs and small quantities of MWCNT bundles. However, even at a low concentration of water ($X = 0.05$), the MWCNT aggregates became visible and were growing in size with increase of $X$. E.g., the mean radius $<r>$ of aggregates (in fact, characterizing the size of their two-dimensional projections) was about 25 μm at $X = 0.05$ and about 850 μm at $X = 1.0$ (in pure water).

The increase of bad solvent (water) concentration $X$ resulted in agglomeration of different small clusters into the bigger ones, and finally, large spanning clusters with the size exceeding the microscope visual field were formed at $X=1$.

Analysis of micro-photos evidenced that the MWCNT clusters were ramified and their surface was tortuous. The satisfactory power law dependences of the number of boxes necessary to cover the boundary of an aggregate $N$ versus the box size $s$ were fulfilled (See, Insert in Fig. 4). Then Eq. (1) was used to estimate the mean fractal dimension $d_f$ of MWCNT aggregate surfaces. Figure 4 shows the example of $d_f$ versus $X$ dependence at fixed concentration of MWCNTs, $C=0.1$ wt. %. The value $d_f$ was close to 1 at very small concentration of water ($X<<1$) and it evidenced that the surface of small clusters was rather smooth. However, the value of $d_f$ increased as $X$ was increasing, and the most intensive increase of $d_f$ was observed at $X\leq0.3$. At high water contents ($X\approx1$), the surface of clusters became highly tortuous and value of $d_f$ was approaching $\approx 1.9$.

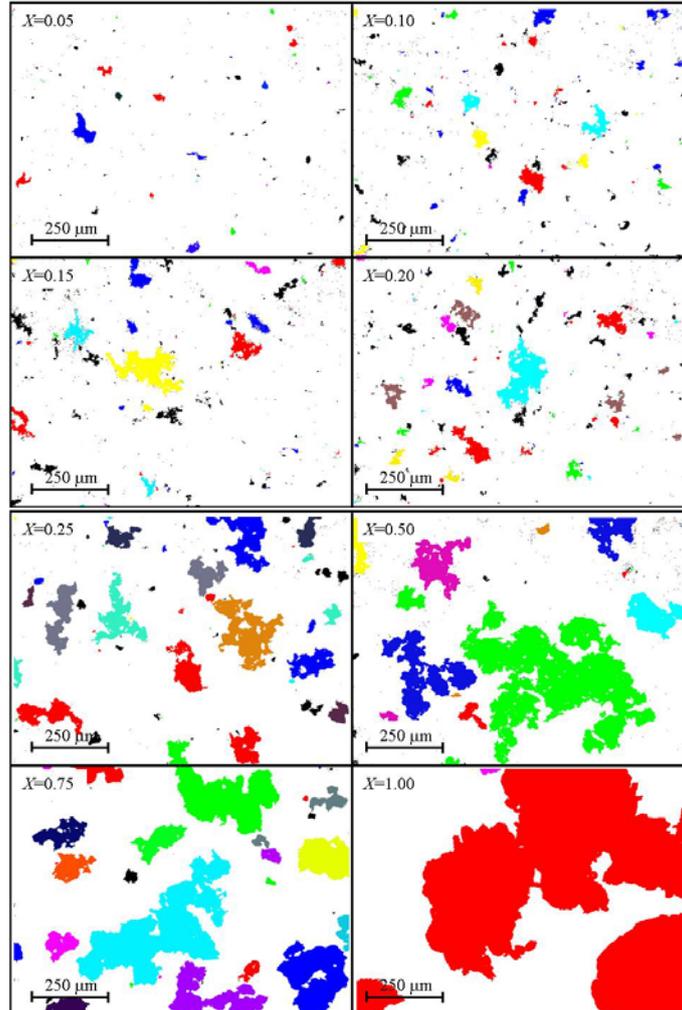

Fig. 3. Micro-photos of MWCNT suspensions at different values of X = [water]/([CHP]+ [water]). Different gray colours correspond to the different clusters. The concentration of MWCNTs was 0.1 wt. % and temperature was 294 K.

Dependencies of the mean radius $<r>$ of aggregates versus concentration of water $X$ at fixed concentration of MWCNTs, $C=0.1$ wt. %, are presented in Fig. 5. Insert shows examples of the integral distribution functions $F(r)$ at different $X$. At small values of $X$ ($0\leq X\leq0.3$), the near-linear $<r>(X)$ dependence was observed. An unessential increase of air–solution surface tension $\gamma$ of CHP + water mixtures was also previously observed in the same concentration range [51] (the data are presented in Fig. 6). The calorimetric data for the heat of CHP mixing with water and behavior of viscosity of CHP + water mixtures at small values of $X$ ($0\leq X\leq0.3$) evidenced the possibility of formation of CHP hydrates in these systems[36]. The CHP hydrates may provoke formation of small MWCNT clusters in MWCNT suspensions as a result of aggregation of individual MWCNTs. The observed unessential aggregation at $0\leq X\leq0.3$ may be explained by high viscosity of CHP + water mixtures (the viscosity of this system goes through the maximum at $X=0.15$[36]) and relatively high dispersing ability of CHP in the presence of hydrates. It is interesting that electrical conductivity of CHP + water mixtures passes through the maximum at $X\approx 0.3$ (Fig. 6) that may also interpreted in favor of formation of CHP hydrates.

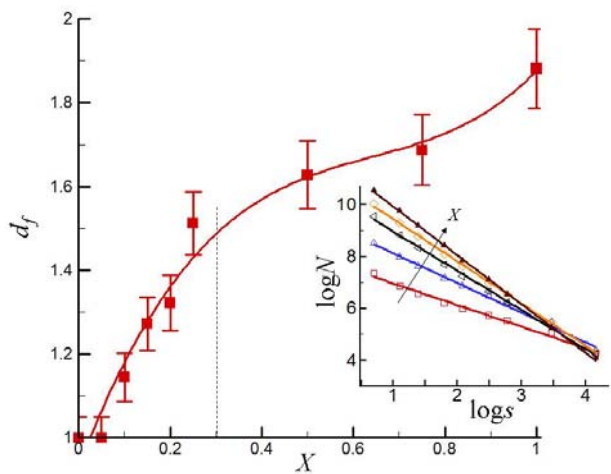

Fig. 4. Mean fractal dimension $d_f$ of MWCNT aggregates versus the value of $X =$ [water]/( [CHP]+ [water]). Insert shows examples of the number of boxes, necessary to cover the boundary of an aggregate $N$, versus the box size $s$. The concentration of MWCNTs was 0.1 wt. % and temperature was 294 K.

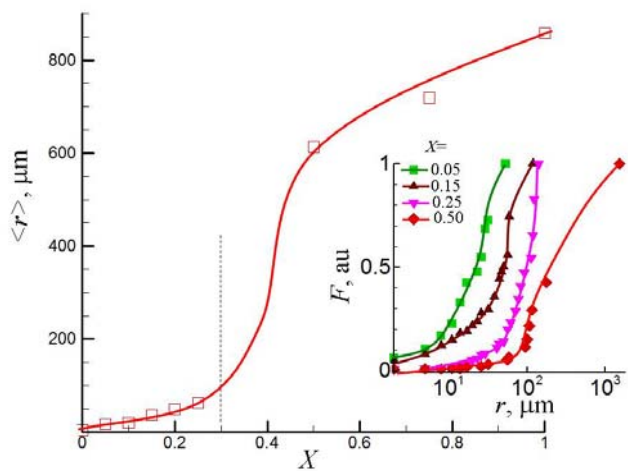

Fig. 5. Mean radius of MWCNT aggregates $<r>$ versus the value of $X =$[water]/( [CHP]+ [water]). Insert shows examples of integral distribution functions $F(r)$ at different $X$. The concentration of MWCNTs was 0.1 wt. % and temperature was 294 K.

At higher concentrations of water ($X>0.3$), the processes of agglomeration of small clusters to larger ones occurred (Fig. 3) and the value of $<r>$ was growing more intensively (Fig. 5). The shape of the distribution function $F(r)$ evidences also that fraction of small clusters ($<r>\leq 20$ μm) and populations of individual MWCNTs became unessential at $X>0.3$ (See, insert in Fig. 5).

It is interesting to note that interval $0.3\leq X\leq 0.9$ corresponds to the plateau on $\gamma(X)$ dependence at the level of $\approx 40$ mJ/m$^2$ (Fig. 6). This plateau reflects the presence of a micellar range in CHP–H$_2$O mixtures and formation of small CHP aggregates containing 20–25 CHP molecules [36]. However, the dispersing ability of CHP + water mixtures at $X>0.3$ was rather bad, though the surface tension of 40 mJ/m$^2$ in this concentration range ($0.3\leq X\leq 0.9$) formally corresponded to the level of good dispersing ability of the solvent [29]. It evidences that the surface tension is not a suitable characteristic for prediction of dispersion ability of MWCNTs in a mixture of good and bad solvents. The observed data, possibly, reflect the differences in the intermolecular forces of CHP and water (e.g., differences in dispersion and hydrogen bonding components of Hansen parameters [35]). The effects of adsorption of surface-active CHP molecules or its micellar aggregates may be also important in determining the dispersing efficiency of MWCNTs in CHP–$H_2O$ mixtures.

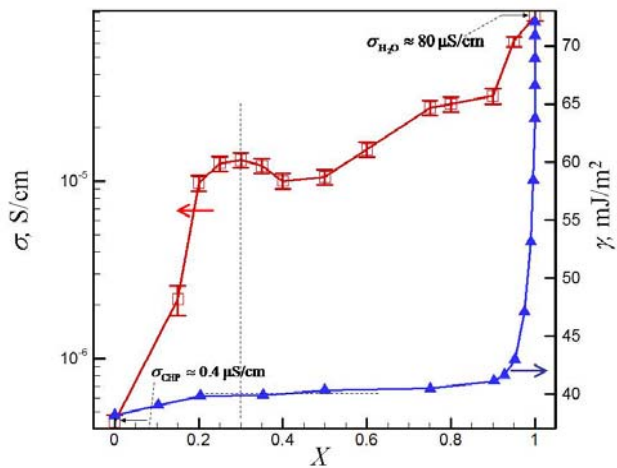

Fig. 6. Electrical conductivity $\sigma$ and surface tension $\gamma$ of CHP+ water mixtures (the data were taken from ref.[36]) versus the value of $X$ =[CHP]/( [CHP]+ [water]). The temperature $T$ was 294 K.

Enhancing of aggregation at high concentrations of water may reflect strengthening of contacts between different MWCNTs. The MWCNTs possess high electrical conductivity and it allows application of electrical conductivity measurements for study of the effects of such strengthening. This technique was previously applied to study of aggregation phenomena of MWCNTs dispersed in different polymers and solvents [39,42]. In order to get a deeper insight into the mechanisms of MWCNT aggregation, more detailed studies of electrical conductivity behaviour of suspensions at different values of $X$ and $C$ were done in this work.

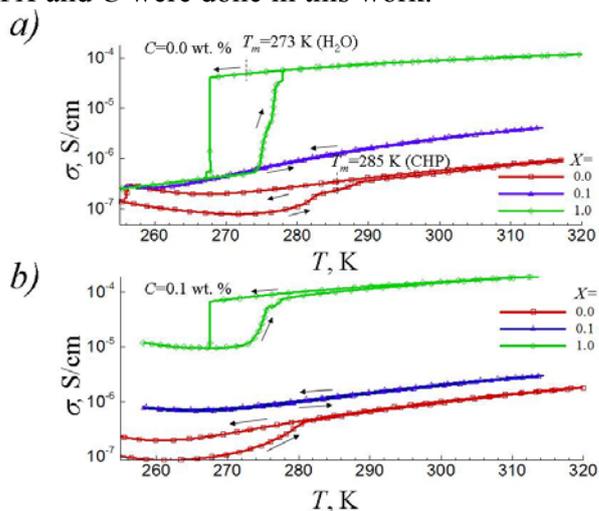

Fig. 7. Temperature dependences of electrical conductivity $\sigma$ in: a) CHP + water mixtures ($C$=0.0 wt. %) and b) MWCNT suspensions in CHP + water mixtures ($C$=0.1 wt. % ) at different values of $X$ =[CHP]/( [CHP]+ [water]). Arrows shows the heating ($\rightarrow$) and cooling ($\leftarrow$) circles.

Figure 7 demonstrates the temperature dependences of electrical conductivity $\sigma$. a) in CHP + water mixtures ($C$=0.0 wt. %) and b) MWCNT suspensions in CHP + water mixtures ($C$=0.1 wt. %) at different values of $X$. In the heating ($\rightarrow$) cycle, the electrical conductivity $\sigma$ was passing through the minimum at $T\approx 275$ K and then grew continuously in pure CHP (Fig. 7a). In the subsequent cooling ($\leftarrow$)cycle, a noticeable hysteresis loop was observed below the melting temperature of CHP, $T_m\approx 285$ K. It evidently reflected the supercooling of CHP. A noticeable hysteresis loop was observed also in temperature dependence of electrical conductivity $\sigma$ in pure water, which reflected retardation of melting in the heating cycle and unessential supercooling of water in the cooling cycle (Fig. 7a) at used scanning rate (2K min$^{-1}$). However, the hysteresis loop was unessential at $X$=0.1, which can be explained by the impact of formation of CHP hydrates on development of the melting and freezing processes in CHP + water mixtures. The similar temperature dependences of electrical conductivity $\sigma$ were observed also in 0.1 wt. % MWCNT suspensions in the CHP + water mixtures (Fig. 7b). Addition of MWCNTs resulted in unessential changes of the shapes of hysteresis loops.

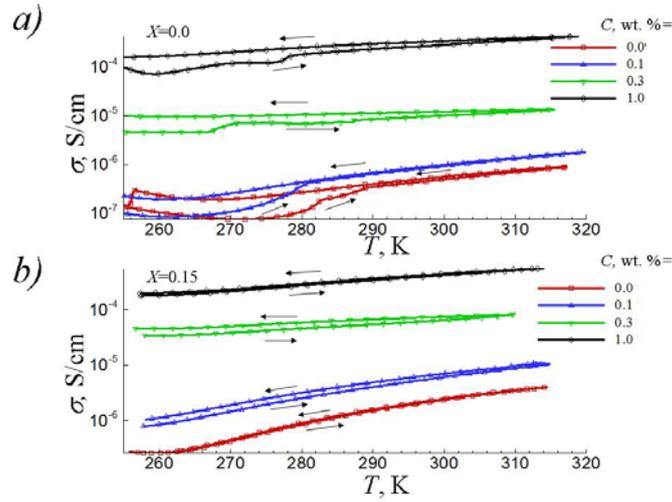

Fig. 8. Temperature dependences of electrical conductivity $\sigma$ in MWCNT suspensions at a) $X$=0 (pure CHP) and b) $X$ =0.15 at different values of $C$. Arrows shows the heating ($\rightarrow$) and cooling ($\leftarrow$) circles.

The effect of concentration of MWCNTs on the temperature dependences of electrical conductivity $\sigma$ for MWCNT suspensions at a) $X$=0 (pure CHP) and b) $X$ =0.15 is presented in Fig. 8. Increase of MWCNT content in the concentration range of 0 wt. % to 1 wt. % caused noticeable increase of electrical conductivity $\sigma$ (by two orders of magnitudes). This increase in the value of $\sigma$ reflected the presence of the percolation transition between non-conductive and conductive states, related with formation of the spanning clusters of MWCNTs. Note that the hysteresis loops were seen more distinctly at $X$=0.0 than at $X$=0.15, which again may be explained by the impact of CHP hydrates. The electrical conductivity data, $\sigma$, obtained within the temperature interval $T = 290 \div 320$ K, followed the Arrhenius law:

$$\sigma \propto \exp(-E/RT) \qquad (3)$$

where $E$ is the activation energy, $R = 8.314$ J mol$^{-1}$ K$^{-1}$ is the universal gas constant.
Figure 9 presents dependences of the activation energy $E$ versus the value of $X$ (at $C = 0$ wt. % and $C = 1$ wt. %) (a) and versus the value of $C$ (at $X = 0$ and $C = 0.15$) (b). The presented data evidence the presence of unessential increase of the activation energy $E$ with increase of $X$ (at constant value of $C$), and sharp decrease of $E$ with increase of $C$ (at constant value of $X$). Note that behavior of the activation energy may reflect the quality of contacts between different MWCNTs. The mechanism of charge transfer in MWCNT loaded suspensions is controlled by the thermally assisted hopping and tunneling of charges between different MWCNTs[39]. The increase or decrease of $E$ value may reflect weakening or strengthening of contacts between MWCNTs, respectively. The electronic conductivity of MWCNTs is characterized by a noticeably smaller temperature coefficient than that of ionic conductivity [43]. So, an increase of the role of charge transport through the MWCNT clusters should essentially decrease the apparent activation energy $E$ in MWCNT+ CHP + water suspensions, and it is clearly seen in Fig. 9b. The similar effects of MWCNT concentration on activation energy were observed also in MWCNT + liquid crystal suspensions [44].

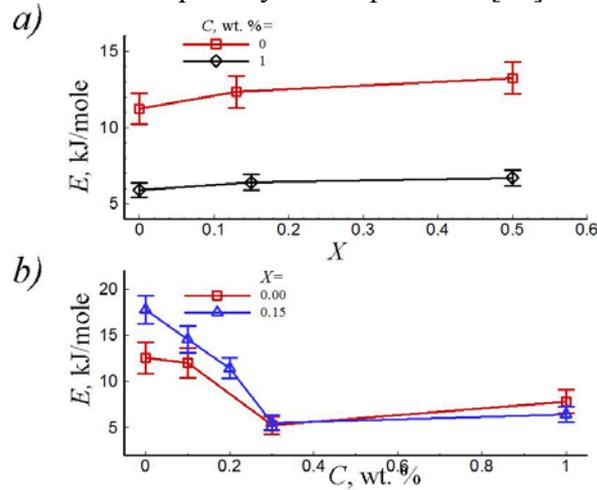

Fig. 9. Activation energy $E$ versus the value of $X$ =[CHP]/( [CHP]+ [water]) at $C$=0 and $C$=1 wt. % (a) and versus the value of $C$ at $X$=0.0 (pure CHP) and $X$=0.15 (b).

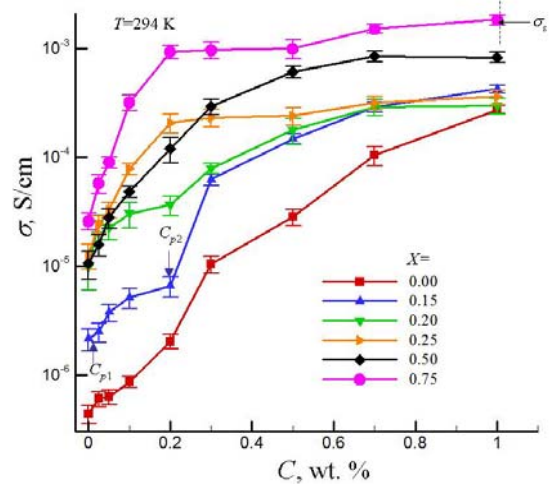

Fig. 10. Electrical conductivity $\sigma$ versus concentration of MWCNTs $C$ at different values of X =[CHP]/( [CHP]+ [water]). The temperature was 294 K.

Figure 10 demonstrates dependencies of electrical conductivity $\sigma$ versus concentration of MWCNTs $C$ at different values of $X$ ($T$=294 K). The shape of all $\sigma(C)$ curves at different values of $X$ evidence the presence of a percolation threshold at rather small concentrations of MWCNTs. E.g., increase in MWCNT concentration within 0.01-0.05 wt. % resulted in the growth of $\sigma$ by several orders of magnitude. The fuzzy type of percolation with multiple thresholds and smeared growth of conductivity was observed practically at all the values of $X$. Note that multiple percolation transitions are typical for composites filled with conductive particles [5,45–49]. Such behavior was previously attributed to segregation of particles [46], the distribution of quality of electrical contacts[47], variations in the shapes and orientations of aggregates and local particle concentration[48], and dependence of the network structure upon the particle concentration [5].

The evident two-step percolation was observed only at small concentrations of water (at $X<0.25$). E.g., two arrows at $C=C_{p1}$ and $C=C_{p2}$ in Fig. 10 show two percolation transitions, identified at $X=0.15$. Such behavior can be explained by formation of the primary percolation network from individual well-dispersed MWCNTs at $C=C_{p1}$ and by the secondary percolation network, composed of the clusters of aggregated MWCNTs at $C=C_{p2}$. These clusters were visually detected on the micro-photos of MWCNT suspensions (Fig. 3). However, it is interesting that the most sharp percolation transition with the smallest percolation threshold was observed in the CHP + water mixtures at rather big content of water that corresponded to commencement of agglomeration of small clusters into larger ones (Fig. 5) (See $\sigma(C)$ curve at $X=0.25$ in Fig.10). Such percolation behavior at $X=0.25$, probably, reflects the process of simultaneous formation of percolation networks consisting of individual MWCNTs and clusters of MWCNTs.

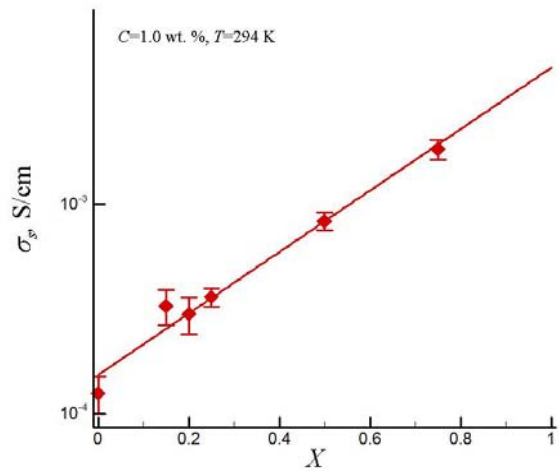

Fig. 11. Saturation value of electrical conductivity $\sigma_s$ (at $C=1.0$ wt. %) versus the value of $X$ =[CHP]/( [CHP]+ [water]). The temperature was 294 K.

At high concentration of MWCNTs ($C\approx 1$ wt. % ), the value of electrical conductivity was reaching the constant of saturation $\sigma\approx\sigma_s$ that was dependent on the value of $X$. Figure 11 presents the $\sigma_s$ versus $X$ dependence at $T=294$ K. It is non-linear and can be approximated by the following exponential equation:
$$\sigma_s=a\exp(bX) \qquad (4)$$
$a=1.67\pm0.12$ mS/cm, $b=3.20\pm0.11$.

Note that this equation predicts that the value of $\sigma_s$ in the bad solvent (water) is approximately 25 times higher that in the good solvent (CHP). Behavior of $\sigma_s(X)$ in mixtures of good and bad solvents may be explained as follows. It is known that the quality of electrical contacts between adjacent MWCNTs is determined by the probability of quantum tunneling between the MWCNTs that decreases exponentially with increase of the distance between MWCNTs [50]. Increase of water contents X results in enhancing of hydrophobic attraction between MWCNTs and in decreasing of distance between adjacent MWCNTs. So, it may be concluded that increase of X improves the electrical contacts between MWCNTs and caused the observed increasing of $\sigma_s$.

## 4. Summary and conclusions

The dispersing quality of MWCNTs in a mixture of good (CHP) and bad (water) solvents may be finely regulated by adjustment of composition of the CHP+ water mixtures. Dispersion of MWCNTs was almost ideal and individual MWCNTs were the dominating population of suspension in the good solvent ($X=0$, pure CHP). Introduction of bad solvent (water) into suspension caused formation of MWCNT aggregates. However, the aggregation was relatively unessential at $0 \leq X \leq 0.3$. It may be explained by high viscosity of CHP + water mixtures and relatively high dispersing ability of CHP + water mixtures in the presence of CHP hydrates. The assumption of CHP hydrates was supported by the observed behavior of electrical conductivity of CHP + water mixtures and also by the previously reported calorimetric data and behavior of viscosity [36]. At higher concentration of water ($X>0.3$), the processes of agglomeration of small clusters into larger ones occurred. The surface of MWCNT clusters was highly tortuous and its fractal dimension $d_f$ increased with increase of $X$, approaching $\approx 1.9$ at $X \rightarrow 1$. The interval $0.3 \leq X \leq 0.9$ corresponds to the micellar range of CHP–$H_2O$ mixtures with approximately constant surface tension of $\gamma \approx 40$ mJ/m$^2$ [36]. Formally this level can indicate good dispersing ability of the solvent [29]. However, the observed intensive aggregation of MWCNTs at $X>0.3$ evidenced that the surface tension is not a suitable characteristic for prediction of dispersion ability in the mixture of good and bad solvents. The studies of electrical conductivity of MWCNT + CHP + water suspensions allowed conclusions regarding the quality of electrical contacts between different MWCNTs in dependence of $X$ and $C$ values. The data evidence the presence of a fuzzy type of percolation with multiple thresholds in the systems under investigation. This behavior was explained by formation of the primary percolation network from individual well-dispersed MWCNTs and by the secondary percolation network composed of the clusters of aggregated MWCNTs. At high concentration of MWCNTs ($C \approx 1$ wt. %), the value of electrical conductivity was reaching the level of saturation at $\sigma \approx \sigma_s$, which was dependent on the value of $X$. Increase of $\sigma_s(X)$ with increase of concentration of the bad solvent $X$ may be explained by enhancement of hydrophobic attraction and decrease of the distance between adjacent MWCNTs. Finally, using of the mixtures of good and bad solvents may be very promising for regulation of aggregation processes in MWCNT suspensions and obtaining of suspensions with controlled aggregation and electro-physical properties.

## Acknowledgment

This study was supported in part by Projects 2.16.1.4, 65/13-H and OKE/10 -13. The authors thank Dr. N.S. Pivovarova for her help with the manuscript preparation.

## References

1. Endo M, Strano MS, Ajayan PM. Potential applications of carbon nanotubes. Topics in Applied Physics. 2008;111:13–62.
2. Hasan T, Sun Z, Wang F, Bonaccorso F, Tan PH, Rozhin AG, et al. Nanotube - polymer composites for ultrafast photonics. Advanced Materials. 2009;21(38-39):3874–99.
3. Pandey G, Thostenson ET. Carbon nanotube-based multifunctional polymer nanocomposites. Polymer Reviews. 2012;52(3-4):355–416.
4. Valentini F, Carbone M, Palleschi G. Carbon nanostructured materials for applications in nano-medicine, cultural heritage, and electrochemical biosensors. Analytical and Bioanalytical Chemistry. 2013;405(2-3):451–65.
5. Kovacs JZ, Mandjarov RE, Blisnjuk T, Prehn K, Sussiek M, Müller J, et al. On the influence of nanotube properties, processing conditions and shear forces on the electrical conductivity of carbon nanotube epoxy composites. Nanotechnology. 2009;20(15):155703.
6. Chen G-X, Li Y, Shimizu H. Ultrahigh-shear processing for the preparation of polymer/carbon nanotube composites. Carbon. 2007;45(12):2334–40.
7. Aguilar JO, Bautista-Quijano JR, Avilés F. Influence of carbon nanotube clustering on the electrical conductivity of polymer composite films. Express Polymer Letters. 2010;4(5):292–9.
8. Grossiord N, Loos J, Regev O, Koning CE. Toolbox for dispersing carbon nanotubes into polymers to get conductive nanocomposites. Chemistry of Materials. 2006;18(5):1089–99.
9. Poorteman M, Traianidis M, Bister G, Cambier F. Colloidal processing, hot pressing and characterisation of electroconductive MWCNT-alumina composites with compositions near the percolation threshold. Journal of the European Ceramic Society. 2009;29(4):669–75.
10. Jiang L, Gao L, Sun J. Production of aqueous colloidal dispersions of carbon nanotubes. Journal of Colloid and Interface Science. 2003;260(1):89–94.
11. Chen Q, Saltiel C, Manickavasagam S, Schadler LS, Siegel RW, Yang H. Aggregation behavior of single-walled carbon nanotubes in dilute aqueous suspension. Journal of Colloid and Interface Science. 2004;280(1):91–7.
12. Lisunova MO, Lebovka NI, Melezhyk O V, Boiko YP. Stability of the aqueous suspensions of nanotubes in the presence of nonionic surfactant. Journal of Colloid and Interface Science. 2006;299(2):740–6.
13. Lamas B, Abreu B, Fonseca A, Martins N, Oliveira M. Assessing colloidal stability of long term MWCNT based nanofluids. Journal of Colloid and Interface Science. 2012;381(1):17–23.
14. Smith B, Wepasnick K, Schrote KE, Cho H-H, Ball WP, Fairbrother DH. Influence of surface oxides on the colloidal stability of multi-walled carbon nanotubes: A structure-property relationship. Langmuir. 2009;25(17):9767–76.
15. Farbod M, Tadavani SK, Kiasat A. Surface oxidation and effect of electric field on dispersion and colloids stability of multiwalled carbon nanotubes. Colloids and Surfaces A: Physicochemical and Engineering Aspects. 2011;384(1-3):685–90.
16. Maeda Y, Kimura S, Hirashima Y, Kanda M, Lian Y, Wakahara T, et al. Dispersion of single-walled carbon nanotube bundles in nonaqueous solution. The Journal of Physical Chemistry B. American Chemical Society; 2004;108(48):18395–7.
17. Bergin SD, Nicolosi V, Cathcart H, Lotya M, Rickard D, Sun Z, et al. Large populations of individual nanotubes in surfactant-based dispersions without the need for ultracentrifugation. Physical Chemistry C. 2008;112(4):972–7.
18. Strano MS, Moore VC, Miller MK, Allen MJ, Haroz EH, Kittrell C, et al. The role of surfactant adsorption during ultrasonication in the dispersion of single-walled carbon nanotubes. Nanoscience and Nanotechnology. 2003;3:81–6.
19. Vaisman L, Wagner HD, Marom G. The role of surfactants in dispersion of carbon nanotubes. Journal of Colloid and Interface Science. 2006;128-130:37–46.
20. Clark MD, Subramanian S, Krishnamoorti R. Understanding surfactant aided aqueous dispersion of multi-walled carbon nanotubes. Journal of Colloid and Interface Science. 2011;354:144–51.
21. Wang B, Han Y, Song K, Zhang T. The use of anionic gum arabic as a dispersant for multi-walled carbon nanotubes in an aqueous solution. Journal of Nanoscience and Nanotechnology. 2012;12(6):4664–9.
22. Loginov M, Lebovka N, Vorobiev E. Laponite assisted dispersion of carbon nanotubes in water. Journal of Colloid and Interface Science. 2012;365(1):127–36.

23. Wang Z, Meng X, Li J, Du X, Li S, Jiang Z, et al. A simple method for preparing carbon nanotubes/clay hybrids in water. Physical Chemistry B. 2009;113:8058–64.
24. Lan YF, Lin JJ. Dispersion of carbon nanocapsules by using highly aspect-ratio clays. Physical Chemistry A. 2009;113:8654.
25. Lan YF, Lin JJ. Clay-assisted dispersion of organic pigments in water. Dyes and Pigments. 2011;90(1):21–7.
26. Furtado CA, Kim UJ, Gutierrez HR, Pan L, Dickey EC, Eklund PC. Debundling and dissolution of single-walled carbon nanotubes in amide solvents. Journal of the American Chemical Society. American Chemical Society; 2004;126(19):6095–105.
27. Gabor T, Aranyi D, Papp K, Karman FH, Kalman E. Dispersibility of Carbon Nanotubes. Materials Science Forum. 2007;537 - 538:161–8.
28. Bergin SD, Sun Z, Rickard D, Streich P V, Hamilton JP, Coleman JN. Multicomponent solubility parameters for single-walled carbon nanotube-solvent mixtures. ACS Nano. 2009;3(8):2340–50.
29. Bergin SD, Sun Z, Streich P, Hamilton J, Coleman JN. New solvents for nanotubes: Approaching the dispersibility of surfactants. Journal of Physical Chemistry C. 2010;114(1):231–7.
30. Coleman JN. Liquid-phase exfoliation of nanotubes and graphene. Advanced Functional Materials. WILEY-VCH Verlag; 2009;19(23):3680–95.
31. Sun Z, O'Connor I, Bergin SD, Coleman JN. Effects of ambient conditions on solvent-nanotube dispersions: exposure to water and temperature variation. Journal of Physical Chemistry C. 2009;113(4):1260–6.
32. Hughes JM, Aherne D, Bergin SD, O'Neill A, Streich P V, Hamilton JP, et al. Using solution thermodynamics to describe the dispersion of rod-like solutes: application to dispersions of carbon nanotubes in organic solvents. Nanotechnology. 2012;23(26):265604.
33. Detriche S, Zorzini G, Colomer JF, Fonseca A, Nagy JB. Application of the Hansen solubility parameters theory to carbon nanotubes. Journal of Nanoscience and Nanotechnology. 2008;8(11):6082–92.
34. Ham HT, Choi YS, Chung IJ. An explanation of dispersion states of single-walled carbon nanotubes in solvents and aqueous surfactant solutions using solubility parameters. Journal of Colloid and Interface Science. 2005;286(1):216–23.
35. Hansen CM. Hansen solubility parameters-A user's handbook. Boca Raton, FL: CRC Press; 2007.
36. Pethica BA, Senak L, Zhu Z, Lou A. Surface and colloidal properties of cyclic amides. 5. N-cyclohexyl-2-pyrrolidone-water mixtures aggregation in solution and adsorption at the air-solution interface. Colloids and Surfaces A: Physicochemical and Engineering Aspects. 2001;186(1-2):113–22.
37. Lou A, Pethica BA, Somasundaran P, Yu X. Phase behavior of N-Alkyl-2-pyrrolidones in aqueous and nonaqueous systems and the effect of additives. Journal of Colloid and Interface Science. 2002;256(1):190–3.
38. Melezhik A V, Sementsov YI, Yanchenko V V. Synthesis of fine carbon nanotubes on coprecipitated metal oxide catalysts. Russian Journal of Applied Chemistry. 2005;78(6):917–23.
39. Lisetski LN, Minenko SS, Ponevchinsky V V, Soskin MS, Goncharuk AI, Lebovka NI. Microstructure and incubation processes in composite liquid crystalline material (5CB) filled with multiwalled carbon nanotubes. Materials Science and Engineering Technology/ Materialwissenschaft und Werkstofftechnik. 2011;42(1):5–14.

40. Hoshen J, Kopelman R. Percolation and cluster distribution. I. Cluster multiple labeling technique and critical concentration algorithm. Phys. Rev. B. American Physical Society; 1976;14(8):3438–45.
41. Liu L, Yang Y, Zhang Y. A study on the electrical conductivity of multi-walled carbon nanotube aqueous solution. Physica. 2004;E24:343–8.
42. Lysenkov E, Lebovka NI, Yakovlev Y V, Klepko V V, Pivovarova NS. Percolation behaviour of polypropylene glycol filled with multiwalled carbon nanotubes and Laponite. Composites Science and Technology. 2012;72(10):1191–5.
43. Eletskii A V. Transport properties of carbon nanotubes. Physics-Uspekhi (Advances in Physical Sciences). 2009;52(3):209–24.
44. Dolgov L, Kovalchuk O, Lebovka N, Tomylko S, Yaroshchuk O. Liquid Crystal Dispersions of Carbon Nanotubes: Dielectric, Electro-Optical and Structural Peculiarities. In: Marulanda JM, editor. Carbon Nanotubes. Vukovar, Croatia: InTech; 2010. p. 451–84.
45. Kovacs JZ, Velagala BS, Schulte K, Bauhofer W. Two percolation thresholds in carbon nanotube epoxy composites. Composites Science and Technology. 2007;67(5):922–8.
46. Zhang MQ, Yu G, Zeng HM, Zhang HB, Hou YH. Two-step percolation in polymer blends filled with carbon black. Macromolecules. 1998;31:6724–6.
47. Calberg C, Blacher S, Gubbels F, Brouers F, Deltour R, Jérôme R. Electrical and dielectric properties of carbon black filled co-continuous two-phase polymer blends. Journal of Physics D: Applied Physics. 1999;32(13):1517–25.
48. Nettelblad B, Martensson E, Onneby C, Gafvert U, Gustafsson A. Two percolation thresholds due to geometrical effects: experimental and simulated results. Journal of Physics D: Applied Physics. 2003;36(4):399–405.
49. McQueen DH, Jager K-M, Pelskova M. Multiple threshold percolation in polymer/filler composites. J. Phys. D: Appl. Phys. 2004;37:2160–9.
50. Rahatekar SS, Shaffer MSP, Elliott JA. Modelling percolation in fibre and sphere mixtures: Routes to more efficient network formation. Composites Science and Technology. 2010;70(2):356–62.